\documentclass[11pt, a4paper]{article}

\usepackage[english]{babel}
\usepackage[utf8]{inputenc}
\usepackage{jcappub_ver}
\usepackage{enumerate}
\usepackage{amsbsy}
\usepackage{amsmath} 
\usepackage{graphics}
\usepackage{mathrsfs}
\usepackage{wrapfig}

\newcommand{\bea}{\begin{eqnarray}} \newcommand{\eea}{\end{eqnarray}}
\newcommand{\el}{\nonumber \\}
\newcommand{\re}[1]{(\ref{#1})}

\newcommand{\pat}{\partial}

\renewcommand{\sec}[1]{section \ref{#1}}
\newcommand{\fig}[1]{figure \ref{#1}}
\newcommand{\brt}[1]{[#1]}
\newcommand{\para}{\paragraph}

\renewcommand{\a}{\alpha}
\renewcommand{\b}{\beta}
\renewcommand{\c}{\gamma}
\renewcommand{\d}{\delta}
\newcommand{\e}{\epsilon}
\renewcommand{\l}{\lambda}

\newcommand{\LCDM}{$\Lambda$CDM\ }

\newcommand{\rmd}{\mathrm{d}}

\newcommand{\nonum}{\\}
\newcommand{\etal} {et al.\ }
\newcommand{\ie}{i.e.\ }

\newcommand{\be}{\bar{e}}
\newcommand{\te}{\tilde{e}}

\newcommand{\thetat}{\tilde{\theta}}
\newcommand{\sigmat}{\tilde{\sigma}}
\newcommand{\htt}{\tilde{h}}
\newcommand{\patl}[1]{\frac{\rmd{#1}}{\rmd\l}}

\newcommand{\kh}{\hat{k}}
\newcommand{\vh}{\hat{v}}

\newcommand{\udot}{\dot{u}}

\newcommand{\Omn}{\Omega_{\mathrm{m0}}}

\newcommand{\sO}{\mathcal{O}}

\newcommand{\cH}{{\cal{H}}}

\newcommand{\PRD}[1]{{\it Phys. Rev.} {\bf D#1}}

\newcommand{\PRL}[1]{{\it Phys. Rev. Lett.} {\bf #1}}

\newcommand{\PLA}[1]{{\it Phys. Lett.} {\bf A#1}}

\newcommand{\MNRAS}[1]{{\it Mon. Not. Roy. Astron. Soc.} {\bf #1}}
\newcommand{\APJ}[1]{{\it Astrophys. J.} {\bf #1}}

\newcommand{\CQG}[1]{{\it Class. Quant. Grav.} {\bf #1}}
\newcommand{\GRG}[1]{{\it Gen. Rel. Grav.} {\bf #1}}
\newcommand{\AaA}[1]{{\it Astron. \& Astrophys.} {\bf #1}}
\newcommand{\PROG}[1]{{\it Prog. Theor. Phys.} {\bf #1}}

\newcommand{\IJMPD}[1]{{\it Int. J. Mod. Phys.} {\bf D#1}}
\newcommand{\PRT}[1]{{\it Phys. Rept.} {\bf #1}}

\title{A covariant treatment of cosmic parallax}

\author{Syksy R\"{a}s\"{a}nen}

\affiliation{University of Helsinki, Department of Physics \\
and Helsinki Institute of Physics \\
P.O. Box 64, FIN-00014 University of Helsinki, Finland}

\emailAdd{syksy {\it dot} rasanen {\it at} iki {\it dot} fi}

\abstract{
The Gaia satellite will soon probe parallax on cosmological distances.
Using the covariant formalism and considering the angle
between a pair of sources, we find parallax for both
spacelike and timelike separation between observation points.
Our analysis includes both intrinsic parallax and parallax
due to observer motion.
We propose a consistency condition that tests the FRW metric
using the parallax distance and the angular diameter distance.
This test is purely kinematic and relies only on geometrical
optics, it is independent of matter
content and its relation to the spacetime geometry.
We study perturbations around the FRW model,
and find that they should be taken into account when
analysing observations to determine the parallax distance.
}

\begin{document}

\maketitle
  
\setcounter{tocdepth}{2}

\setcounter{secnumdepth}{3}

\section{Introduction} \label{sec:intro}

\para{Measuring cosmic parallax.}

Parallax, \ie change in the angular position of objects on
the sky when viewed from different locations, has had important
implications for cosmology from the debate on the heliocentric
system in the 16th century to the modern distance ladder.
Parallax is an attractive way to measure distances, 
because in contrast to angular diameter or luminosity,
no information about source properties is needed, as the reference
quantity is the distance between observation points, which is directly
measured. However, so far distance measurements using parallax have been restricted
to nearby objects in the Galaxy. The longest parallax distance currently
measured, by the Hipparcos satellite\footnote{http://sci.esa.int/hipparcos},
is of the order 100 pc.

The possibility of measuring parallax over cosmological distances,
cosmic parallax, was first raised in 1935 by McCrea \cite{Mccrea:1935},
though he considered it impossible in practice.
If only the difference in the position of the Earth with respect to the
Sun is considered, the maximum spatial separation between observation
points, called the baseline, is two astronomical units (AU).
In the 1970s it was suggested that advances in spacecraft technology
would make cosmic parallax measurements feasible in the near future
\cite{Weinberg:1970, Weinberg:1972, Kardashev:1973, Novikov:1977, Novikov:1978}.
In 1986 Kardash\"{e}v proposed using the distance travelled by the Earth
with respect to the cosmic microwave background (CMB)
as a baseline, instead of the difference in position relative to
the Sun \cite{Kardashev:1986}.
The Earth moves with respect to the CMB with a velocity of
369 km/s (with an annual modulation of 30 km/s due to motion around the
Sun) \cite{Aghanim:2013}\footnote{Assuming that the almost all of the
CMB dipole is due to a Doppler effect. Without this assumption, the
velocity is determined to be (384 $\pm$ 139) km/s \cite{Aghanim:2013}.},
so the change in position during one year is 78 AU.
Furthermore, this baseline increase is secular, unlike that due to Earth's
motion around the Sun.
Because primordial cosmological perturbations are close to adiabatic
\cite{Planckinflation}, the average rest frame of distant sources,
\ie the frame of statistical homogeneity and isotropy of the matter
distribution, is close to the rest frame of the CMB.

Cosmic parallax due to motion with respect to the CMB will soon
be probed for the first time by the Gaia
satellite\footnote{http://www.cosmos.esa.int/web/gaia},
which was launched on December 19, 2013. Over a period of five years,
Gaia is expected to measure the angular positions of about 500 000 quasars
and 3 million galaxies at cosmological distances with
a precision of order 100 $\mu$as
\cite{Mignard:2005, Lindegren:2008, Slezak:2007, Ding:2009, Quartin:2009, Quercellini:2010, Robin:2012, Gaiasp}.

\para{Cosmic parallax as a test of homogeneity and isotropy.}

The angular diameter distance $D_A$ and the luminosity
distance $D_L$ are related trivially by the Etherington relation
$D_L=(1+z)^2 D_A$ \cite{Etherington:1933, Ellis:1971}, but the
parallax distance $D_P$ contains independent information.
McCrea pointed out already in 1935 that this makes it possible to test
the FRW metric by comparing measurements of $D_P$ and
measurements of $D_A$ (or $D_L$) \cite{Mccrea:1935}. In \cite{Weinberg:1970}
the more limited point was made that if the universe is described by
the FRW metric, measuring both $D_P$ and $D_L$ makes it possible
to determine the metric fully, \ie to find both the
scale factor $a(t)$ and the curvature constant $K$.
The parallax distance was given for an arbitrary spacetime
with zero null shear in \cite{Jordan:1961} and properly derived
in \cite{Rosquist:1988}, and derived for the general case in \cite{Kasai:1988}.
Like the original work of McCrea, these papers studied
the angular position of a single source relative to a constant
baseline direction
between two observation points separated by a spacelike interval.
We refer to this setup as the 'classic' parallax case.

The classic analysis is not straightforwardly applicable to
real observations, because in them the interval between
observation points is timelike, not spacelike. Also, in practice positions
of sources on the sky are measured relative to each other,
instead of comparing a single source to a constant
baseline direction. In 1988, Hasse and Perlick derived
the general condition under which relative positions of
sources on the sky remain constant, assuming
that the world lines of observers and sources are integral
curves of the same timelike vector field
\cite{Hasse:1988} (see also \cite{Perlick:1990}).
In this case, parallax of all pairs of sources vanishes for
all observers if and only if the vector field is conformally
stationary. (In the case of observers comoving with dust matter,
this means that the spacetime is either FRW or stationary or both.)
This is the same as the condition for all observers to measure vanishing
CMB anisotropy \cite{Perlick:1990, EGS, Clarkson:2000}.
The CMB conditions are stable in one direction, but not in the other.
If the spacetime is close to FRW, the CMB is almost isotropic.
However, small CMB anisotropy does not imply that the
spacetime would be close to FRW (or stationary), even when the matter is dust
\cite{almostEGS, moreEGS, Rasanen:2009a, Maartens:2011}.
There has been no similar analysis of the stability of the parallax conditions.
If the universe is close to FRW, is parallax necessarily small?
If parallax is small for all observers, does that imply that the universe is
close to FRW? Also, the analysis in \cite{Hasse:1988, Perlick:1990} does
not consider a difference between the velocity field of observers
and the velocity field of sources, which is central in the
parallax due to our motion with respect to the frame of
homogeneity and isotropy.

Using the parallax to test for deviations from the FRW metric has come up
recently in connection with models where a large spherical inhomogeneity
has been suggested as an alternative to dark energy
\cite{Quercellini:2008, Quartin:2009, Quercellini:2010, Amendola:2013},
as well as proposed models with large-scale anisotropy
\cite{anisopara, Quercellini:2010, Amendola:2013}.
The idea is that if the universe is on
large scales well described by the FRW metric, the parallax is small,
so any detection indicates deviation from the FRW case.
The parallax due to our motion with respect to the frame of
statistical homogeneity and isotropy has been considered
noise, and cosmic parallax has been presented as being due to
anisotropic expansion, \ie shear, and characterised as a property
of spacetime, as opposed to parallax due to observer motion.
However, while all models with shear have non-vanishing cosmic
parallax, there are shear-free models with cosmic parallax
\cite{Krasinski:2012}.
In any case, shear is a property of a velocity field, and the
spacetime geometry constrains the velocity field, but does not determine it.
The separation between parallax due to our motion and parallax
as a property of spacetime depends on the choice of frame.
It has also been proposed that instead of treating the parallax
due to our motion as noise, it could be used to put constraints
on dark energy models, assuming that the universe
is described by the FRW metric \cite{Ding:2009}.
These studies are model-specific, and have all used classic
parallax results based on spacelike separation.
However, it is possible to use the parallax distance for
a kinematic test of the FRW metric that relies only
on geometrical optics, and is independent of the matter content
or the dynamics that relates it to the spacetime geometry
(usually given by the Einstein equation).

We will expand and complement previous work.
In \sec{sec:gen} we derive the parallax for a pair of sources in a
general spacetime in terms of covariant quantities. We consider
both spacelike and timelike separation, and include
both the parallax due to deviation of the spacetime from conformal stationarity
and the parallax due to difference between observer and source velocity.
We also discuss the general relation between $D_P$ and $D_A$.
In \sec{sec:FRW} we specialise to the FRW universe. We introduce
a consistency condition between $D_P$ and $D_A$ that is specific to
the FRW metric.
In \sec{sec:pert} we consider the perturbed FRW case and study
the stability of the FRW results.
In \sec{sec:disc} we discuss non-perturbative deviations from the
FRW case, different kinds of tests of homogeneity and isotropy
and the accuracy with which the parallax distance will be probed by Gaia.
In \sec{sec:conc} we summarise our results and mention open issues.

\section{General spacetime} \label{sec:gen}

\subsection{Geometrical setup}

\para{The two frames.}

We mostly follow the notation of \cite{Rasanen:2009b};
for reviews of the covariant approach, see
\cite{Ehlers:1961, Ellis:1971, Ellis:1998c, Clarkson:2000, Tsagas:2007}.
We consider two timelike velocity fields, $n$ and $u$.
Both are normalised to unity, $n \cdot n=u \cdot u=-1$, but are otherwise
arbitrary (we use the notation $a\cdot b\equiv g_{\a\b} a^\a b^\b$, where
$g_{\a\b}$ is the metric, for any vectors $a$ and $b$).
We take $u$ to be the velocity of the observer, and we will later
take $n$ to correspond to the frame of statistical homogeneity
and isotropy. Without loss of generality, we can write $u$ in
terms of $n$ and a vector $v$ that is orthogonal to $n$,
\bea \label{n}
  u^\a = \gamma ( n^\a + v^\a ) \ ,
\eea

\noindent where $\gamma\equiv -n\cdot u=(1-|v|^2)^{-1/2}$, with
$|v|^2\equiv v\cdot v$ and $v\cdot n=0$.
The tensors that project on the rest spaces orthogonal to $n$ and $u$ are
\bea \label{h}
  h_{\a\b} &\equiv& g_{\a\b} + n_\a n_\b \el
  h_{\a\b}^{(u)} &\equiv& g_{\a\b} + u_\a u_\b \ .
\eea

The derivative of $u$ can be decomposed as
\bea \label{ugrad}
  \nabla_\b u_\a
  &=& \frac{1}{3} h^{(u)}_{\a\b} \theta^{(u)} + \sigma^{(u)}_{\a\b} + \omega^{(u)}_{\a\b} - \udot_\a u_\b \ ,
\eea

\noindent where $\theta^{(u)}\equiv\nabla_\a u^\a$ is the volume expansion rate,
$\sigma_{\a\b}^{(u)}\equiv h^{(u)}_{\a\c} h^{(u)}_{\b\d} \nabla^\d u^\c-\frac{1}{3}\theta^{(u)} h^{(u)}_{\a\b}$
is the shear tensor,
$\omega^{(u)}_{\a\b} \equiv \nabla_{[\b} u_{\a]} + \udot_{[\a}u_{\b]}$
is the vorticity tensor, and
$\udot^\a\equiv u^\b\nabla_\b u^\a$ is the acceleration vector.
Overdot refers to derivative along $u$.

\para{Light bundles.}

In the geometrical optics approximation light travels on null geodesics
\cite{Schneider:1992} (page 93).
We denote the momentum of a light ray labelled $A$ by $k_A$,
and we have $k_A\cdot k_A=0$ and $k_A^\a \nabla_\a k_A=0$.
Photon energy measured in the two frames is
\bea \label{E}
  E_A^{(n)} &=& - n \cdot k_A \el
  E_A^{(u)} &=& - u \cdot k_A \ .  
\eea

\noindent The photon momentum can be decomposed as
\bea \label{kdec}
  k_A^\a &=& E_A^{(n)} ( n^\a + e_A^\a ) \el
  &=& E_A^{(u)} ( u^\a + r_A^\a ) \ ,
\eea

\noindent with $n\cdot e_A=0$, $e_A\cdot e_A=1$ and
$u \cdot r_A=0$, $r_A \cdot r_A=1$.
We introduce the 'normalised' dimensionless photon momentum
\bea \label{khat}
  \kh_A^\a &\equiv& {E_A^{(u)}}^{-1} k_A^\a = u^\a + r_A^\a \ .
\eea

\begin{wrapfigure}[15]{r}[0pt]{0.35\textwidth}
\scalebox{0.5}{\includegraphics[angle=0, clip=true, trim=0.0cm 0.0cm 0.0cm 0.0cm]{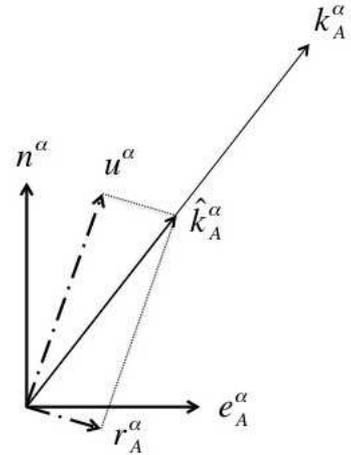}}
\caption{The relation between $k_A$, $\kh_A$, $n$, $e_A$, $u$ and $r_A$.}
\label{fig:vectors}
\end{wrapfigure}

\noindent The relation between the two decompositions of
$k_A$ is illustrated in \fig{fig:vectors}.

We define a tensor that projects orthogonally to $k_A$ as
\bea \label{htt}
  \htt_{A\a\b} &=& g_{\a\b} + n_{\a} n_{\b} - e_{A\a} e_{A\b} \el
  &=& g_{\a\b} - {E_A^{(n)}}^{-2} k_{A\a} k_{A\b} + 2 {E_A^{(n)}}^{-1} k_{A(\a} n_{\b)} \ .
\eea

\noindent Note that $\htt_{A\a\b}$ has been chosen to project orthogonally
also to $n$ (and thus $e_A$), whereas $\kh_A$ is defined with respect to
$u$. The observer frame is more relevant for observations,
but $n$, once it is chosen to correspond to the frame of statistical 
homogeneity and isotropy, is better adapted to the spacetime geometry.

The derivative of $k_A$ can be decomposed as
\bea \label{kgrad}
  \nabla_\b k_{A\a} &=& \frac{1}{2} \htt_{A\a\b} \thetat_A + \sigmat_{A\a\b} + k_{A(\a} P_{A\b)} \ ,
\eea

\noindent where
$\thetat_A\equiv\htt^{\a}_{A\b} \nabla_\a k_A^\b=\nabla_\a k_A^\a$
is the expansion rate of the area of a bundle of null geodesics and
$\sigmat_{A\a\b}\equiv \htt_{A\a}^{\d} \htt_{A\b}^{\c} \nabla_\c k_{A\d} - \frac{1}{2} \htt_{A\a\b} \thetat_A$
is the null shear. We have $\sigmat_{A\a\b} k_A^\b=0$, $P_A\cdot k_A=0$.
The null shear scalar is defined as
$\sigmat_A\equiv\sqrt{\frac{1}{2}\sigmat_{A\a\b}\sigmat_A^{\a\b}}$.
The scalars $\thetat_{A}$ and $\sigmat_A$ do not depend on the choice
of $\htt_{A\a\b}$, as long as it is orthogonal to $k_A$.
The vector $P_A$ does depend on this choice; for the tensor \re{htt} we have
\bea \label{P}
  P_A^\a = - 2 {E_A^{(n)}}^{-1} n^\b \nabla_\b k_A^\a - {E_A^{(n)}}^{-2} n^\b n^\c \nabla_\c k_{A\b} k_A^\a \ .
\eea

\noindent We also define $P_{\perp A}\equiv P_A + u\cdot P_A \kh_A$,
which is orthogonal to both $k_A$ and $u$.

\subsection{Spacelike separation} \label{sec:genspace}

\para{Parallax.}

We consider two sources that send bundles of light rays to
the observer and label them $A=1$ and $A=2$. The observer sees
the sources in the directions $-r_1$ and $-r_2$, so the angle
$\varphi$ separating them is given by
\bea \label{g}
  g \equiv \cos\varphi &=& r_1 \cdot r_2 = \kh_1 \cdot \kh_2 + 1 \ .
\eea

\begin{figure}
\begin{minipage}[t]{7.7cm} 
\scalebox{1.0}{\includegraphics[angle=0, clip=true, trim=0cm 0cm 0cm 0cm, width=\textwidth]{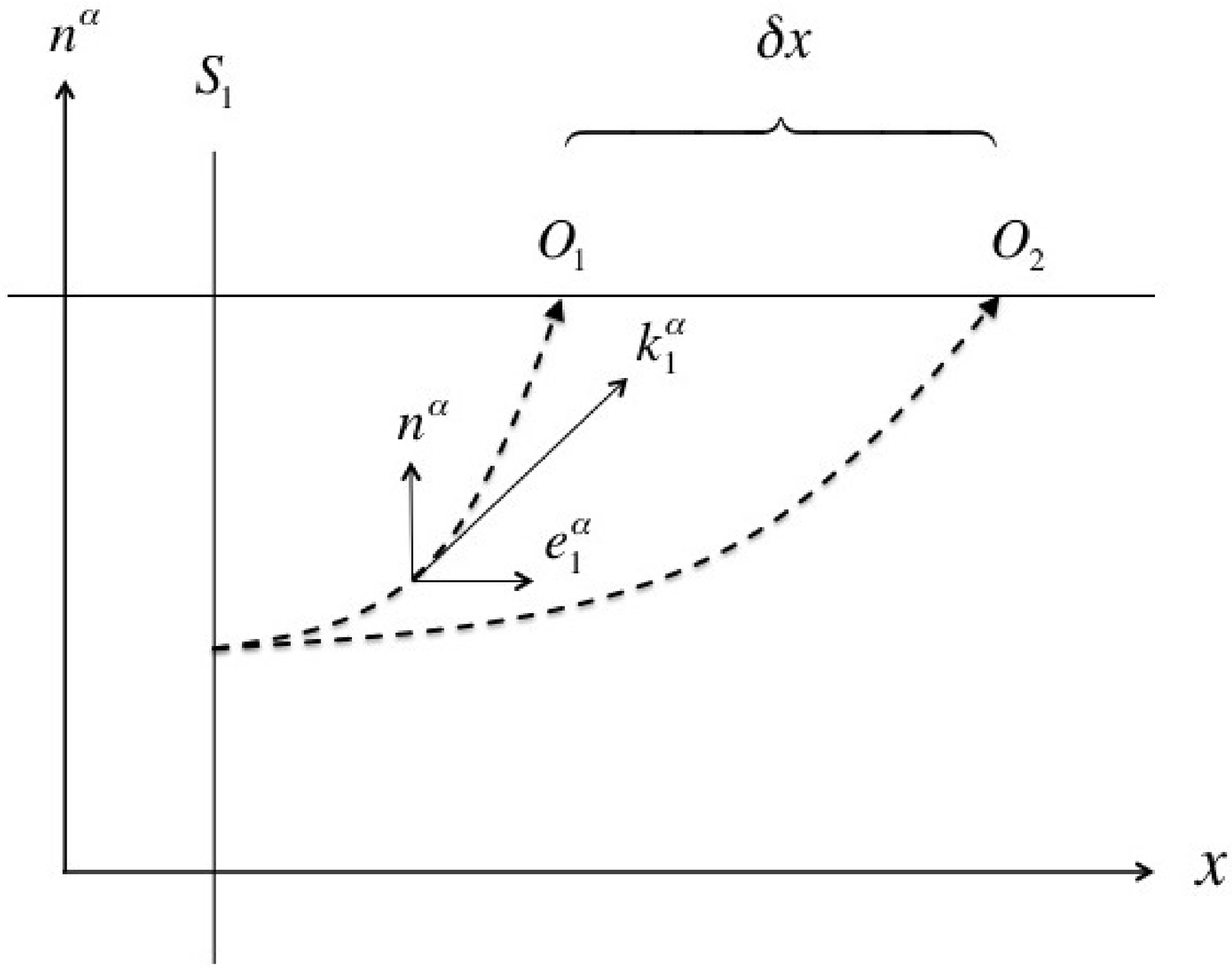}}
\begin{center} {\bf a)} \end{center}
\end{minipage}
\begin{minipage}[t]{7.7cm}
\scalebox{1.0}{\includegraphics[angle=0, clip=true, trim=0cm 0cm 0cm 0cm, width=\textwidth]{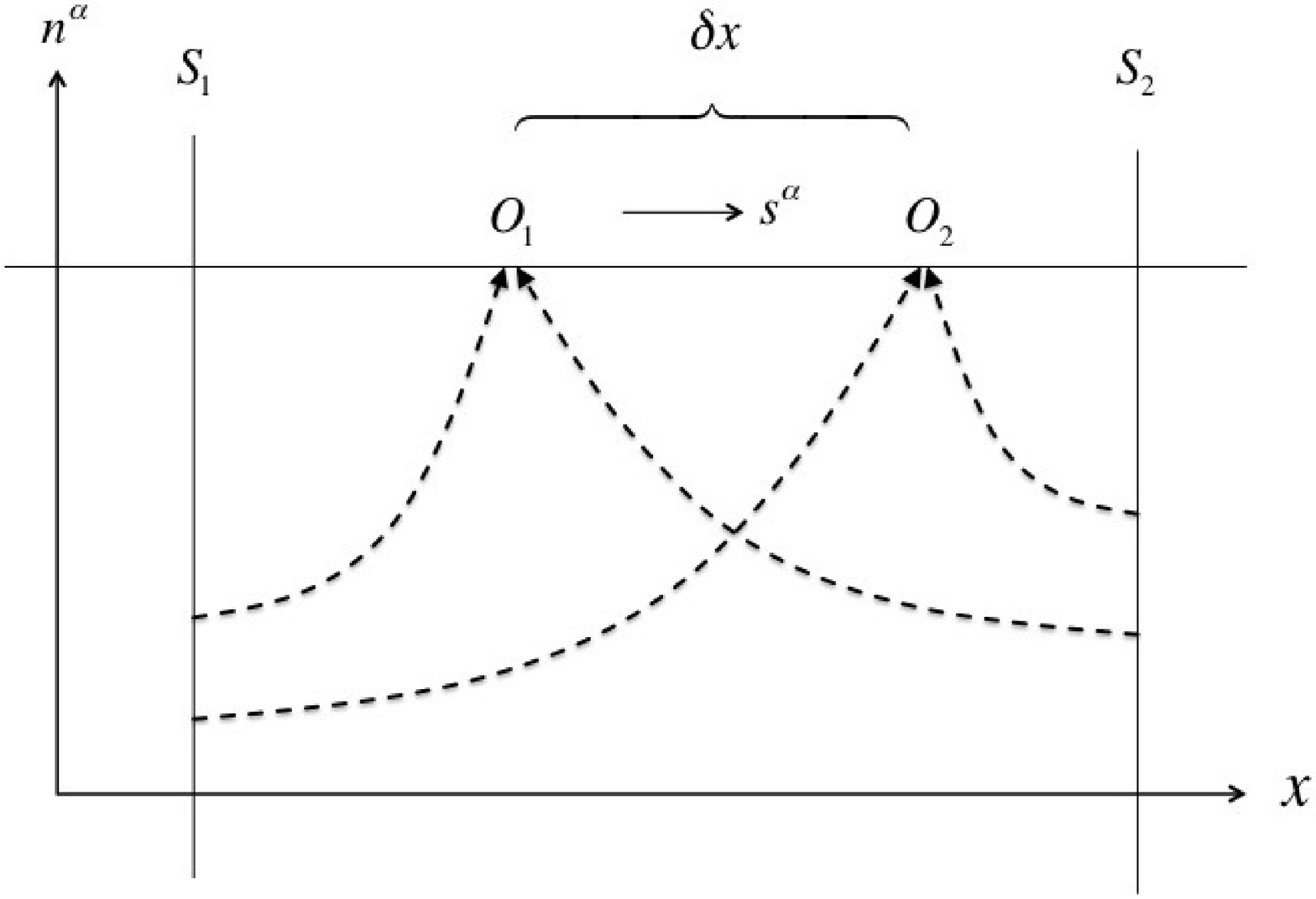}}
\begin{center} {\bf b)} \end{center}
\end{minipage}
\begin{minipage}[t]{7.7cm}
\vspace{0.5cm}
\scalebox{1.0}{\includegraphics[angle=0, clip=true, trim=0cm 0cm 0cm 0cm, width=\textwidth]{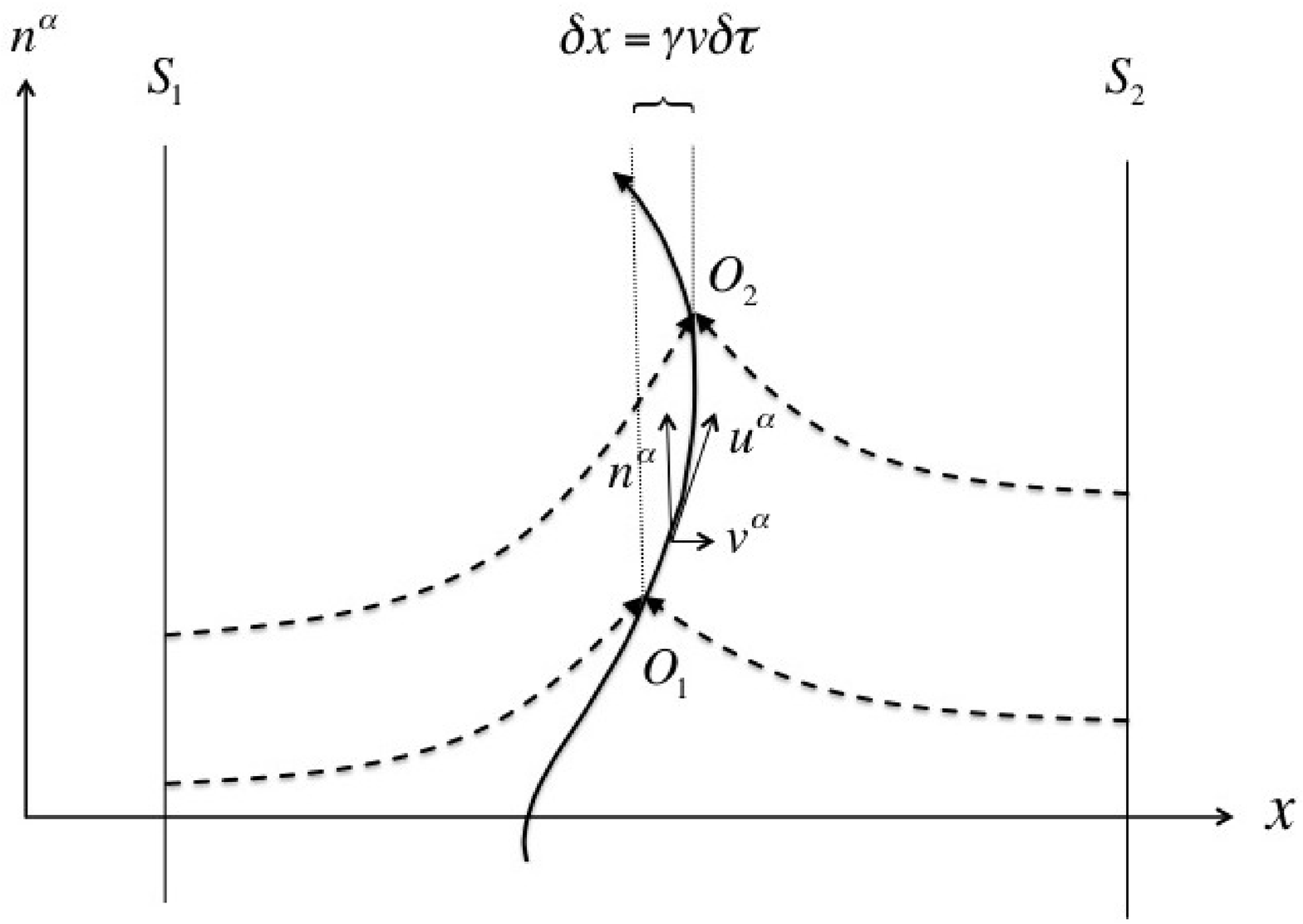}}
\begin{center} {\bf c)} \end{center}
\end{minipage}
\caption{Schematic illustration of the different parallax cases.
a) Classic parallax. One source $S_1$ is observed at points $O_1$ and $O_2$
separated by the spacelike interval $\delta x$ on the hypersurface
orthogonal to $n$.
b) Two sources $S_A$ are observed at points $O_1$ and $O_2$ separated
by the spacelike interval $\delta x$ in the direction $s$ on the hypersurface
orthogonal to $n$.
c) Two sources $S_A$ are observed at points $O_1$ and $O_2$ separated
by the timelike interval $\delta\tau$ in the direction $u$.
The projected spatial distance between the observation points on the
hypersurface orthogonal to $n$ is $\delta x=\gamma v \delta\tau$.}
\label{fig:parallax}
\end{figure}

\noindent Following \cite{Hasse:1988}, we define parallax as the
change of $g$. According to this definition, rigid rotation of the
celestial sphere relative to a local inertial frame, as in the
G\"{o}del universe, is not parallax. The case of classic parallax
with one source and spacelike separation between observation points,
as well as our case with two sources, for both spacelike
and timelike separation, is illustrated in \fig{fig:parallax}.

We first consider observation points separated by a spacelike interval.
We denote the spacelike unit vector in the direction of separation by
$s$, so $s\cdot s=1$, and we also assume that
$s\cdot n=0$.\footnote{Were $s$ to contain a component
parallel to $n$, we would get a linear combination of
\re{gprime} and the $v$-independent part of \re{gdot}.} We obtain
\bea \label{gprime}
  g' &\equiv& s^\a \nabla_\a g \el
  &=& {E_1^{(u)}}^{-1} \kh_2^\b s^\a \nabla_\a k_{1\b} + \left( {E_1^{(u)}}^{-1} u^\b s^\a \nabla_\a k_{1\b} + \kh^\b_1 s^\a \nabla_\a u_\b \right) \kh_1  \cdot \kh_2 + ( 1 \leftrightarrow 2 ) \el
  &\simeq& \frac{1}{2} r_1\cdot s \ P_{\perp 1}\cdot r_2 + ( r_2\cdot s - r_1\cdot r_2 \ r_1\cdot s ) \left( \frac{1}{2} {E_1^{(n)}}^{-1} \thetat_1 - \frac{1}{3} \theta^{(u)} \right) + {E_1^{(n)}}^{-1} \sigmat_{1\a\b} r_2^\a s^\b \el
 && - ( 1 - r_1\cdot r_2 ) \left( \sigma_{\a\b}^{(u)} r_1^\a s^\b + \omega_{\a\b}^{(u)} r_1^\a s^\b \right) + ( 1 \leftrightarrow 2 ) \ ,
\eea


\noindent where we have used the decompositions \re{ugrad} and \re{kgrad},
$( 1 \leftrightarrow 2 )$ refers to the same expression as the one
written down, but with the labels 1 and 2 interchanged, and in the last
equality we have taken the
limit $|v|\ll1$ and dropped all terms that contain any factors of $v$
without derivatives.
Even if $|v|\ll1$, it is possible that $\nabla_\b v_\a\sim\nabla_\b n_\a$,
so the expansion rate $\theta^{(u)}$ measured by the observer in the
$u$ frame can be very different from the expansion rate in the
$n$ frame. This is the case for realistic observers located
in gravitationally bound structures, where the difference in the velocity
with respect to the frame of statistical homogeneity and isotropy
is non-relativistic, but the relative difference in the expansion rate is unity.

If the observer is displaced a small distance $\delta x$ in the direction
$s$, the change in $g$ is $\delta g=g' \delta x$, so the change in
the angle is $\delta\varphi=-(\sin\varphi)^{-1} g' \delta x$. We thus have
\bea \label{phiperxspace}
  \frac{\delta\varphi}{\delta x} &=& - \frac{1}{\sin\varphi} g' = - \frac{1}{\sqrt{1- ( r_1\cdot r_2 )^2 }} g' \ ,
\eea

\noindent where all quantities are evaluated at the position of the observer.

\para{Comparison to the classic case.}

The result \re{gprime}, \re{phiperxspace} is different from the classic
parallax result for spacelike separation
\cite{Jordan:1961, Kasai:1988, Rosquist:1988}. The reason for
the difference is most transparent when comparing to the analysis of
\cite{Kasai:1988}, where the light ray from the second observation point
is transported to the first observation point. We
transport the observer instead, and as a result pick up terms involving
the gradient of $u$ in addition to the gradient of $k_A$.
(There is also an overall sign difference relative to
\cite{Kasai:1988, Rosquist:1988}, because we consider
future-oriented rays instead of past-oriented rays.)
The classic parallax involves the change of angular position of
one source with respect to a constant baseline direction.
We can get closer to that setup by choosing
$-r_2$ to point in the direction of the source, $-r_2=s$.
The term involving $\thetat_2$ in \re{gprime} then vanishes.
In classic considerations of parallax, the baseline is also
(to zeroth approximation) orthogonal to the line of sight,
so we take $r_1\cdot s=0$. We then have
\bea \label{paraspace}
  \frac{\delta\varphi}{\delta x} &\simeq& \frac{1}{2} P_{\perp 2}\cdot r_1 + \frac{1}{2} {E_1^{(n)}}^{-1} \thetat_1 - \frac{1}{3} \theta^{(u)} + {E_1^{(n)}}^{-1} \sigmat_{1\a\b} s^\a s^\b - {E_2^{(n)}}^{-1} \sigmat_{2\a\b} r_1^\a s^\b \el
  && + \sigma_{\a\b}^{(u)} r_1^\a s^\b + \omega_{\a\b}^{(u)} r_1^\a s^\b - \sigma_{\a\b}^{(u)} s^\a s^\b \ .
\eea

\noindent We can get even closer to the classic analysis, where
the baseline is considered to be constant, by neglecting $P_2$ and
$\sigmat_{2\a\b}$, which describe changes in null geodesic bundle 2
related to the baseline direction, to obtain
\bea \label{paraspace2}
  \frac{\delta\varphi}{\delta x} &\simeq& \frac{1}{2} {E_1^{(n)}}^{-1} \thetat_1 + {E_1^{(n)}}^{-1} \sigmat_{1\a\b} s^\a s^\b - \frac{1}{3} \theta^{(u)} + \sigma_{\a\b}^{(u)} r_1^\a s^\b + \omega_{\a\b}^{(u)} r_1^\a s^\b - \sigma_{\a\b}^{(u)} s^\a s^\b \ .
\eea

\noindent If we consider an observer whose velocity
field has vanishing expansion rate, shear and vorticity, only the
first two terms of \re{paraspace2} remain, and we recover
the classic parallax result \cite{Jordan:1961, Rosquist:1988, Kasai:1988}.
However, this does not describe a realistic cosmological situation,
because even if the observer is located in a stabilised region
with $\theta^{(u)}=0$, the shear and vorticity are in general
not zero. This could induce significant angular asymmetry in the
parallax, as shear and vorticity can be of the same order of magnitude
as the average expansion rate, which in turn is generally
of the same order of magnitude as ${E_1^{(n)}}^{-1}\thetat_1$.
(This will be clear when we consider the FRW
spacetime in \sec{sec:FRW}.)
However, the situation with spacelike separation
is not relevant for real observations of cosmic parallax.
Let us now consider the realistic case where the
observation points are separated by a timelike interval.

\subsection{Timelike separation} \label{sec:gentime}

\para{Parallax.}

The observationally relevant case is the one where the
separation between the observation points is timelike,
as they are on the worldline of an observer moving along
a timeline curve tangent to $u$. The rate of change of the angle
$\varphi$ along the curve is given by
\bea \label{gdot}
  \dot g &\equiv& u^\a \nabla_\a g \el
  &=& {E_1^{(u)}}^{-1}  \kh_2^\b u^\a \nabla_\a k_{1\b} +  \left( {E_1^{(u)}}^{-1} u^\b u^\a \nabla_\a k_{1\b} + \udot\cdot\kh_1 \right) \kh_1\cdot\kh_2 + ( 1 \leftrightarrow 2 ) \el
  &=& - \frac{1}{2} P_{\perp 1}\cdot \kh_2 + \kh_1\cdot\kh_2 \ \udot\cdot\kh_1 + \gamma \kh_2^\a v^\b \left( \frac{1}{2} \htt_{1\a\b} {E_1^{(n)}}^{-1} \thetat_1 + {E_1^{(n)}}^{-1} \sigmat_{1\a\b} \right) \el
  && + \gamma^2 v^\a v^\b {E_1^{(u)}}^{-1} \left( \frac{1}{2} \htt_{1\a\b} \thetat_1 + \sigmat_{1\a\b} \right) \kh_1\cdot\kh_2 + ( 1 \leftrightarrow 2 ) \el
   &\simeq& - \frac{1}{2} P_{\perp 1}\cdot r_2 - ( 1- r_1\cdot r_2 ) \udot\cdot r_1 \el
  && + (r_2\cdot v - r_1\cdot r_2 \ r_1\cdot v) \frac{1}{2} {E_1^{(n)}}^{-1} \thetat_1 + {E_1^{(n)}}^{-1} \sigmat_{1\a\b} r_2^\a v^\b + ( 1 \leftrightarrow 2 ) \ ,
\eea

\noindent where we have again applied the decomposition \re{kgrad}
and in the last equality taken the limit $|v|\ll1$ and dropped terms
that contain more than one power of $v$ (in this case there are no
derivatives of $v$).
All quantities are again evaluated at the location of the observer.
We have not expressed $\udot$ in terms of $n$ frame quantities,
as we will consider geodesic observers, $\udot=0$.
Note that we have kept terms proportional to $v$, in contrast to
the spacelike separation case \re{gprime}, because in the timelike case
motion with respect to the $n$ frame is important.

If $v=0$ (this can be thought of as decomposing
$\nabla_\b k_{A\a}$ with respect to the observer velocity), we have
\bea
  \dot g &=& - \frac{1}{2} P_{\perp 1}\cdot r_2 - ( 1 - r_1\cdot r_2) \udot\cdot r_1 + ( 1 \leftrightarrow 2 ) \ ,
\eea

\noindent and the contributions of $\thetat_A$ and $\sigmat_{A\a\b}$
disappear. If we assume that also the sources move along
curves tangent to $n=u$, all parallaxes vanish for
all observers if and only if $\sigma^{(u)}_{\a\b}=0$ and
$\nabla_{[\b}(\udot_{\a]}-\frac{1}{3}\theta^{(u)} u_{\a]})=0$ \cite{Hasse:1988}.

Let us now assume that $v\neq0$ and find the change of angle
relative to the $n$ frame spatial displacement.
We denote the proper time measured by the observer by $\tau$.
During a small time interval $\delta\tau$, the observer
has, in the $n$ frame, moved the spatial distance
$\delta x=\gamma |v|\delta\tau$, and the angle has changed by
$\delta\varphi=-(\sin\varphi)^{-1} \dot g \delta\tau$. We thus have
\bea \label{phiperxtime}
  \frac{\delta\varphi}{\delta x} = - \frac{1}{\sin\varphi} \frac{\dot g}{\gamma |v|} = - \frac{1}{\sqrt{ 1- ( r_1\cdot r_2 )^2 }} \frac{\dot g}{\gamma |v|} \ ,
\eea

\noindent where all quantities are again evaluated at the position
of the observer.

\para{Comparison to the classic case and the spacelike case.}

As in the case of spacelike separation, we get closer to
the classic situation by considering angular position with respect to
a constant baseline, so we take $-r_2=\vh$, with $\vh\equiv v/|v|$,
and $r_1\cdot r_2=0$. We then have from \re{gdot} and \re{phiperxtime}
\bea \label{paratime}
  \frac{\delta\varphi}{\delta x} &\simeq& \frac{1}{|v|} \left[ - \frac{1}{2} P_{\perp 1}\cdot\vh + \frac{1}{2} P_{\perp 2}\cdot r_1 + \udot\cdot r_1 - \udot\cdot\vh \right] \el
  && + \frac{1}{2} {E_1^{(n)}}^{-1} \thetat_1 + {E_1^{(n)}}^{-1} \sigmat_{1\a\b} \vh^\a \vh^\b - {E_2^{(n)}}^{-1} \sigmat_{2\a\b} r_1^\a \vh^\b \ ,
\eea

\noindent The first line in \re{paratime} is the
parallax that in general remains even in the case $v=0$, and it is
enhanced by $1/|v|$ relative to the terms on the second line,
which vanish when $v=0$.
This can be understood as dividing the parallax into a contribution
due to deviation of $n$ from conformal stationarity and a contribution
due to observer motion with respect to the $n$ frame.
If $n$ is identified with the frame of statistical homogeneity and
isotropy, the former contribution can be called 'intrinsic'
and considered to be due to the properties of the spacetime,
and the latter can be thought of as being due to observer
motion in space. Such a separation is frame-dependent.

For timelike separation, the change of angle
$\delta\varphi$ between two points separated by spatial
distance $\delta x$ is different than in the case where the observer
has moved only spatially. This is to be expected, because in the case of
timelike separation also the sources have moved, not just the observer.
The difference is twofold: in the timelike case, terms related
to the local expansion rate, anisotropic shear and vorticity are absent,
while in the spacelike case, there are no terms that persist even in the
case of zero spatial motion.

If observer motion is geodesic,
the time derivative of $v$ does not enter into \re{gdot}
or \re{phiperxtime}. However, in writing $\delta x=\gamma |v|\delta\tau$
we have implicitly assumed that the time interval $\delta\tau$
is short enough that $v$ does not change appreciably.
For realistic observations, this is not quite true, as
the velocity with respect to the CMB has a 10\% annual
modulation  due to motion around the Sun, which should
be taken into account in a more detailed analysis.

In \cite{Ding:2009}, it was argued that when considering the
parallax due to our motion with respect to the frame
of statistical homogeneity and isotropy, the expansion term
appearing in \re{paraspace2} should be added to the parallax,
using the FRW background expansion rate.
In other words, we should consider a baseline that
expands in time instead of a fixed baseline.
In most of the literature, the fixed baseline has been used,
with the exception of \cite{Weinberg:1970, Weinberg:1972}.
From \re{paratime} we see that the timelike case corresponds
to the fixed baseline equation, and no expansion
term needs to be added, unlike argued in \cite{Ding:2009}.
However, the classic result gives only part of the parallax,
and there are terms both due to the spacetime
geometry and observer motion.
We will look at the quantitative importance of these terms
in the case of perturbed FRW spacetime in \sec{sec:pert}, but
let us first consider the definition of the parallax distance,
and its relation to angular diameter distance, given that
the general result \re{gdot} and \re{phiperxtime} contains more terms
than are present in the classic definition of parallax distance.

\subsection{Parallax distance and angular diameter distance} \label{sec:DPDA}

\para{Definition of parallax distance.}

The classic parallax distance is defined as
$D_P=(\delta\varphi/\delta x)^{-1}$, with the observation
points separated by a spacelike interval and the baseline
kept constant and orthogonal to the direction of observation.
If we were to neglect the local expansion rate, shear and vorticity,
\re{paraspace} would give the result
$\delta\varphi/\delta x = \frac{1}{2} {E^{(n)}}^{-1} \thetat + {E^{(n)}}^{-1} \sigmat_{\a\b} s^\a s^\b$
(in this subsection, we consider only a single source,
so we drop the subscript $A$).
This agrees with the classic result
\cite{Jordan:1961, Rosquist:1988, Kasai:1988}.
However, for real observations there are extra terms,
including ones that are non-zero for $\delta x=0$.
The classic definition of $D_P$ also has the shortcoming that when
considering pairs of sources, the parallax varies with direction.
We prefer to have a simple definition of the parallax distance
to a source, one that is determined from $(\delta\varphi/\delta x)^{-1}$
but is not equal to it. We define, as in \cite{Ellis:1971},
\bea \label{DP}
  D_P^{-1} \equiv \frac{1}{2} {E^{(n)}}^{-1} \thetat \ ,
\eea

\noindent where we consider a light bundle that converges
at the source, and the expression is evaluated at the observer.
We have defined $D_P$ in the $n$ frame. As $\thetat$
is frame-independent, in the observer frame we have
$D_P^{(u)}\equiv E^{(u)}/E^{(n)} D_P=\gamma (1-v\cdot e) D_P$.

\para{Relation to angular diameter distance.}

An important property of the parallax distance, noted already
in \cite{Mccrea:1935}, is that it is not trivially related to
the angular diameter distance $D_A$, unlike the luminosity distance $D_L$.
In general relativity the Etherington relation $D_L=(1+z)^2 D_A$
always holds \cite{Etherington:1933, Ellis:1971}, so measuring the luminosity
distance as a function of redshift does not provide extra information
compared to the angular diameter distance. The relation between the
parallax  distance and the angular diameter distance is more subtle.
Both distances are defined using $\thetat$, the area expansion
rate of the light bundle. However, in the case of $D_A$, light rays
converge at the observer, whereas for $D_P$ they converge at the source.
The evolution of $\thetat$ is governed by the Sachs equations,
\bea \label{Sachs}
  \patl\thetat + \frac{1}{2} \thetat^2 + 2 \sigmat^2 &=& - R_{\a\b} k^\a k^\b \el
  \htt_{\a}^{\ \ \c} \htt_{\b}^{\ \ \d} \patl{\sigmat_{\c\d}} + \thetat \sigmat_{\a\b} &=& - k^\mu k^\nu \htt_{\a}^{\ \ \c} \htt_{\b}^{\ \ \d} C_{\mu\gamma\nu\d} \ ,
\eea

\noindent where $\l$ is the affine parameter so that $\patl{}=k^\a\nabla_\a$,
$R_{\a\b}$ is the Ricci tensor and $C_{\a\b\c\d}$ is the Weyl tensor.
For a light ray converging at the observer, the initial
condition is $\thetat_0=-\infty$, whereas for a
light ray converging at the source, it is $\thetat_{\mathrm{s}}=\infty$;
the subscript 0 refers to quantities evaluated at the
observer's location in time and space, and s refers to quantities
evaluated at the source. In the former case, we have
$D_A\propto \exp \left( {\frac{1}{2} \int\rmd\l \thetat} \right)$,
whereas in the latter case we have instead
$D\propto \exp \left( {\frac{1}{2} \int\rmd\l_0 \thetat} \right)$,
where $D\equiv (1+z) D_A$ and $\l_0$ is the affine parameter
at the observer \cite{Ellis:1971, Rosquist:1988}\footnote{The
distance $D$ is known as galaxy area distance \cite{Ellis:1971},
proper motion distance \cite{Weinberg:1972},
transverse comoving distance \cite{Hogg:1999},
photon count distance \cite{Cattoen:2007}
and, most commonly, comoving angular diameter distance.}.
The parallax distance is directly related to the local value of $\thetat$,
because it involves the difference between two light rays at the
observer's location, whereas $D$ and $D_A$ depend on $\thetat$ via an integral.

Let us consider the case with negligible null shear.
(In the real universe, the null shear is small for typical
light rays \cite{Munshi:2006, Clarkson:2011c}.)
By comparing the two solutions of \re{Sachs} corresponding
to different initial conditions, with rays either
converging at the source or at the observer,
and using the definition \re{DP}, we get the
following relation between $D_P$ and $D_A$ \cite{Rosquist:1988}:
\bea \label{DPDA}
  {E_0^{(n)}}^{-1} \frac{\rmd D_P^{-1}}{\rmd\l} = \frac{1}{D_A^2} \quad\Leftrightarrow\quad D_P^{-1} = \int\rmd\l E_0^{(n)} \frac{1}{D_A^2} \ .
\eea

\noindent If we instead consider the change with respect to $\l_0$,
we have \cite{Rosquist:1988}
\bea \label{DPD}
  D_P^{-1} = {E_0^{(n)}}^{-1} \frac{1}{D} \frac{\rmd D}{\rmd\l_0} \ .
\eea

\noindent The relation \re{DPDA} is a general consistency condition
akin to the Etherington relation. It relates two distances measured
at the observer, and it holds as long geometrical optics is valid
(in parallax case, null shear also has to be negligible).
Unlike in the case of the Etherington
relation, expressing \re{DPDA} in terms of observational quantities
requires converting from the affine parameter to the redshift;
we discuss this in more detail in the FRW case in \sec{sec:FRW}.
The relation \re{DPD} involves the difference in the distance $D$ to the
same source as observed from different positions, so measuring it
on cosmological scales is not feasible in the foreseeable future.

\section{FRW spacetime} \label{sec:FRW}

\subsection{Metric, frame and light bundles} \label{sec:geom}

Let us consider parallax in the FRW spacetime, with the metric
\bea \label{metricFRW}
  \rmd s^2 = - \rmd t^2 + a(t)^2 \left( \frac{1}{1 - K R^2} \rmd R^2 + R^2 \rmd\Omega^2 \right) \ ,
\eea

\noindent where the spatial curvature parameter $K$ is a constant.
We normalise the scale factor to unity today, $a(t_0)=1$.

Taking $n$ to correspond to the frame where the universe
looks homogeneous and isotropic, we have for the observer frame
\bea \label{FRWtheta}
  \theta^{(u)} &\simeq& 3 H + \pat_i v^i + \frac{1}{2} u^\a \nabla_\a |v|^2 \ ,
\eea

\noindent where $H\equiv\frac{1}{a}\frac{\rmd a}{\rmd t}$.
and we have taken into account $|v|\ll1$. Note
that derivatives of $v$ can nevertheless be comparable to, or
even larger than, background quantities. For an observer
on a Solar orbit, $|v|\sim 10^{-3}$. However, $|v|$ changes by about 10\%
on a timescale of one year, so the last term in \re{FRWtheta} is
$\sim 10^{-7}$ yr$^{-1}\sim 10^3 H_0$.
(The contribution from motion around the center of the Galaxy
is much smaller: $|v|$ changes by order unity on a timescale
of $10^8$ years, so we have $\sim 10^{-14}$ yr$^{-1}\sim 10^{-4} H_0$.)
In the FRW spacetime, such an observer trajectory is non-geodesic,
because $\udot=0$ implies $u^\a \nabla_\a |v|^2=-2H |v|^2$.
Even if we neglected small-scale motion around the Sun
or the center of the Galaxy, this evolution would be problematic,
because for an observer in a stabilised region like a galaxy,
the expansion rate should be vanish, but \re{FRWtheta} is not
zero, because $3H$ and $\pat_i v^i$ in general have different time-dependence.
However, if we opt for non-geodesic motion to get the desired
timelike curve, there will be fictitious non-gravitational forces,
whose effect on the parallax may be
different than those of the real gravitational interactions that
are responsible for the small-scale motion and zero expansion rate.
These problems arise because deviations from homogeneity and
isotropy that cause the observer trajectory to deviate from the
background are not included in the description.
The resolution is to consistently consider deviations from
FRW spacetime. We will look at linear perturbations in the next
section, though we will see that it is not possible to
treat stabilised regions in linear perturbation theory.
In any case, as we saw in \sec{sec:gen}, the local expansion rate
(and the shear and the vorticity) only affect the parallax in the
case of spacelike separation, whereas it is the case of
timelike separation that is observationally relevant.

For the light bundle quantities we have
\bea \label{FRWkgrad}
  \frac{1}{2} {E_A^{(n)}}^{-1} \thetat_A &=& H + \frac{\sqrt{1 - K R^2}}{a R} \ ,\quad \sigmat_{A\a\b} = 0 \ , \quad P_A^\a = H {E_A^{(n)}}^{-1} k_A^\a \ ,
\eea

\noindent where the source is located at $R=0$; recall that
$D_P^{-1} =\frac{1}{2} {E_{A0}^{(n)}}^{-1} \thetat_{A0}$ .
From $P_A||k_A$ it follows that $P_{\perp A}=0$.
Before discussing $D_P$, let us consider parallax in
the cases of spacelike and timelike separation.

\subsection{Spacelike separation} \label{sec:spaceFRW}

For observation points separated by a spacelike interval, we obtain
from \re{gprime}, \re{phiperxspace} and \re{FRWkgrad}
\bea \label{FRWspace}
  \frac{\delta\varphi}{\delta x} &\simeq& - \frac{r_2\cdot s - r_1\cdot r_2 \ r_1\cdot s}{\sqrt{1- ( r_1\cdot r_2 )^2 }} \left( \frac{1}{2} {E_1^{(n)}}^{-1} \thetat_1 - \frac{1}{3} \theta^{(u)} \right) \el
  && + \frac{1 - r_1\cdot r_2}{\sqrt{1- ( r_1\cdot r_2 )^2 }} \left( \sigma_{\a\b}^{(u)} r_1^\a s^\b + \omega_{\a\b}^{(u)} r_1^\a s^\b \right) \ .
\eea


\noindent For an observer comoving with the frame of
homogeneity and isotropy ($v=0$), we have
\bea \label{FRWspacevzero}
  \frac{\delta\varphi}{\delta x} &=& - \frac{r_2\cdot s - r_1\cdot r_2 \ r_1\cdot s}{\sqrt{1- ( r_1\cdot r_2 )^2 }} ( D_{P1}^{-1} - H ) + ( 1 \leftrightarrow 2 ) \ ,
\eea

\noindent where $D_{PA}$ is parallax distance to source $A$.
The parallax between a pair of sources involves the parallax distances
to both of them. If we take $-r_2=s$ and $r_1\cdot r_2=0$,
as in \sec{sec:gen}, we get the classic parallax result
for a single source for expanding baseline,
$\frac{\delta\varphi}{\delta x}=D_{P1}^{-1}-H$ \cite{Weinberg:1970, Weinberg:1972}.
If we instead consider an observer whose space is not expanding,
$\theta^{(u)}=0$, and also put the shear and vorticity to zero,
we get the classic result for a fixed baseline,
$\frac{\delta\varphi}{\delta x}=D_{P1}^{-1}$ \cite{Mccrea:1935}.

\subsection{Timelike separation} \label{sec:timeFRW}

When the observation points are along the worldline of an observer
moving with velocity $u$, we have from \re{gdot} and \re{FRWkgrad}
\bea \label{gdotFRW}
  \dot g &\simeq& - ( 1 - r_1 \cdot r_2 ) \udot\cdot r_1 + ( r_2 \cdot v - r_1 \cdot r_2 \ r_1 \cdot v ) D_{P1}^{-1} + ( 1 \leftrightarrow 2 ) \ .
\eea

\noindent All terms in \re{gdotFRW} vanish for $v=0$
(as the FRW frame vector $n$ is geodesic,
$\udot$ goes to zero when $v\rightarrow0$).
For observers whose rest frame coincides with the frame of
homogeneity and isotropy, relative angular positions on the sky
remain constant in time, as is obvious from the symmetry of the spacetime.
However, other observers, with $v\neq0$, measure parallax given by
\bea \label{FRWtime}
  \frac{\delta\varphi}{\delta x} &\simeq& \frac{1 - r_1 \cdot r_2}{\sqrt{1 - ( r_1\cdot r_2 )^2}} \frac{1}{|v|} \udot\cdot r_1 - \frac{r_2\cdot\vh - r_1\cdot r_2 \ r_1\cdot\vh}{\sqrt{ 1 - ( r_1\cdot r_2 )^2 }} D_{P1}^{-1} + ( 1 \leftrightarrow 2 ) \ .
\eea

\noindent For a geodesic observer, the first term is zero and we get the
same result as in the case of spacelike separation, \re{FRWspace},
but without the terms related to the expansion rate, shear and
vorticity. In the FRW spacetime, observations of parallax
give a straightforward measure of the parallax distance. Before
considering the effect of deviations from the FRW case, let us
introduce a FRW consistency condition involving $D_P$ and $D$.

\subsection{FRW consistency condition} \label{sec:con}

Let us return to the relation between the parallax
distance $D_P$ and the distance $D$
(or equivalently $D_A$ or $D_L$) discussed in \sec{sec:DPDA}.
From \re{FRWkgrad} we have the parallax distance in the
FRW case, taking into account $a_0=1$ and $R_0=D$ \cite{Mccrea:1935},
\bea \label{dPFRW}
  d_P &=& \frac{d}{ d + \sqrt{1-k d^2} } \ ,
\eea

\noindent where we have defined
$d_P\equiv H_0 D_P$, $d\equiv H_0 D$, $k\equiv K/H_0$.
Note that the parallax distance would asymptote to a finite value even
if $d$ were to grow without limit --  though in realistic cosmological
models, $d$ approaches a finite value of order unity at large $z$.
In other words, the change in the angular position of
faraway objects approaches a constant of the order $H_0 \delta x$
instead of becoming arbitrarily small, as in Euclidean space. This means
that the observational difficulty due to the small change in
angular separation saturates quickly with increasing redshift,
though the problem of observing distant sources remains,
as the luminosity distance does grow without limit at large $z$.

We can solve for $k$ from \re{dPFRW} to obtain
\bea \label{kP}
  k_P &=& \frac{1}{d^2} - \left( \frac{1}{d_P} - 1 \right)^2 \ ,
\eea

\noindent where we have added the subscript $P$ to indicate that
the spatial curvature constant has been solved from the relation between
$d_P$ and $d$.
The relation \re{kP} provides a test of the FRW metric. If
$d_P$ and $d$ are measured independently and the combination
$k_P$ is not constant, we can conclude that light propagation
in the universe is not described by the FRW metric.
(The reverse does not hold true: $k_P$ may be constant even if
light propagation cannot be described by the FRW metric.)
Such a test was already pointed out in
\cite{Mccrea:1935}, and it is possible because spatial
curvature affects $d_P$ and $d$ differently.
Note that this test is purely kinematic. It is based on the geometrical
optics treatment of light propagation, and it is independent of the
matter content of the universe or the equation of motion that relates 
the matter to the spacetime geometry.

In the FRW spacetime, $d$ is related to the expansion rate by
\bea \label{dFRW}
  d(z) = \frac{1}{\sqrt{-k}} \sinh\left(\sqrt{-k}\int_0^z \frac{\rmd \tilde z}{h(\tilde z)} \right)
\eea

\noindent for any value of $k$, where we have defined $h\equiv H/H_0$ and assumed that
$H>0$ (except possibly at isolated points), and $z$ is the redshift
measured by an observer comoving with matter.
From this we can solve for $k$ to obtain \cite{Clarkson:2007b}
\bea \label{kH}
  k_H &=& \frac{1 - h^2 d'^2}{d^2} \ ,
\eea

\noindent where the subscript $H$ refers to the fact that the spatial
curvature constant has been solved from the relation between $h$
and $d$, and prime denotes derivative with respect to $z$.
The relation \re{kH} was presented in \cite{Clarkson:2007b}
as way to test of the FRW metric by independently measuring the
distance $d$ and the expansion rate $h$.

Combining \re{dPFRW} and \re{kH} to eliminate $k$, we have
\bea \label{DPDH}
  d_P = \frac{d}{ d + h d' } \ .
\eea

\noindent The relation \re{DPDH} is a FRW consistency condition
that does not involve the constant $k$, but instead relates the
three functions $d_P$, $d$ and $h$.

The consistency condition \re{dPFRW} can be seen as another
test of the FRW relation \re{dFRW} between the distance $d$
and the expansion rate $h$. The relation \re{DPDA} discussed in
\sec{sec:DPDA} gives $d_P(\l)$, if we know $d_A(\l)$, for any
spacetime (assuming zero null shear). However, going from the affine
parameter $\l$ to the observable redshift $z$ involves the
expansion rate (and in general also the shear and the acceleration
vector). Inserting the FRW result $E_0^{(n)}\rmd\l=-\rmd z/[H (1+z)^2]$
into the general relation \re{DPDA} between $d_P$ and $d_A$ and
using the FRW result \re{dFRW} between $d$ and $h$ gives the
FRW relation \re{dPFRW} between $d_P$ and $d$, which can also
be written as
\bea \label{dPh}
  d_P = \frac{1}{ 1 + \sqrt{-k} \coth\left( \sqrt{-k} \int_0^z \frac{\rmd\tilde z}{h(\tilde z)} \right) } \ .
\eea

\noindent We can treat \re{dPh} as another consistency condition,
involving $d_P$ and $h$.
However, \re{kP} has the virtue that it involves only the distances
$d$ and $d_P$, no integrals or derivatives, unlike  \re{kH} or \re{dPh}.
Also, determining $d_P$ does not require any information
about source properties, whereas finding the expansion rate $h$
as a function of redshift involves assumptions about the sources,
whether using large-scale structure statistics  (in particular
baryon acoustic oscillations) or galaxy ages.

If light propagation on large scales is not described by the
four-dimensional FRW metric, then neither $k_P$ nor $k_H$
will be constant in general.
Possible deviations that could lead to violation of the FRW relation
between $d$ and $h$ and thus to non-constant $k_P$ and $k_H$
include the universe not being statistically homogeneous and
isotropic on large scales \cite{February:2009},
the existence of more than three spatial dimensions \cite{Ferrer} and
the average behaviour on large scales not being FRW despite
statistical homogeneity and isotropy, \ie significant backreaction
\cite{Ellis:2005, Rasanen:2011a, Buchert:2011, Rasanen:2008b, Rasanen:2009b, Boehm:2013, Lavinto:2013}.
But the real universe is in any case not exactly FRW.
Before applying the consistency condition \re{kP} as a test of whether
the FRW metric correctly describes the mean optical properties of
spacetime, we should know the corrections due to perturbations around the FRW case.

\section{Perturbed FRW spacetime} \label{sec:pert}

\subsection{Perturbation theory}

Let us consider a general perturbed FRW spacetime. We write the metric as
(we follow the notation of \cite{Malik:2008}; however, we do not assume
that the background is spatially flat)
\bea \label{metricpert}
  \rmd s^2 = a(\eta)^2 \left[ - ( 1 + 2 \phi ) \rmd \eta^2 + 2 B_i \rmd \eta \rmd x^i + \left( \gamma_{ij} + 2 C_{ij} \right) \rmd x^i \rmd x^j \right] \ ,
\eea

\noindent where $\gamma_{ij}$ is the metric of three-dimensional
homogeneous and isotropic space with constant curvature $6 K$.
Indices of the metric perturbations are raised and lowered with $\gamma_{ij}$.
We denote background quantities by overbar and perturbations by $\delta$.

We assume that metric perturbations are $\sO(\e)$, and
that their derivatives with respect to $t$ are $\sO(H\e)$.
We also assume that spatial derivatives increase the order of
magnitude, and that $\e\ll\pat\e\ll1$, where $\pat$ stands
for $\pat_i/(a|H|)$ (see \cite{Rasanen:2011b} for a more careful
accounting of the smallness conditions). We also assume that
$\delta u^i,|v|\sim\sO(\pat\e)$. We do not assume that
terms of order $\sO(\pat^2\e)$ are small.

Under the above assumptions, corrections to the redshift
are small, and the average expansion rate is also close
to the FRW value \cite{Rasanen:2011b}. However,
corrections to the luminosity distance can be of order unity
\cite{Enqvist:2009}. If the universe is statistically homogeneous
and isotropic, and the distribution evolves slowly compared to
the time it takes light to travel the homogeneity scale, then
it is expected that the luminosity distance is close to the FRW
case, because the average expansion rate is near the FRW value
\cite{Rasanen:2008b, Rasanen:2009b} (see also \cite{Lavinto:2013}).
We are interested in the correction to the parallax
$\frac{\delta\varphi}{\delta x}$ and the parallax distance $d_P$
(and hence the FRW consistency condition \re{kP}).

We only consider the observationally relevant case of timelike
separation and geodesic observers. The parallax is then determined
in terms of $P_{\perp A}$, $\thetat_A$ and $\sigmat_{A\a\b}$ by
\re{gdot} and \re{phiperxtime}. In the exact FRW case,
$P_{\perp A}=0$. However, in the perturbed case the intrinsic
parallax does not vanish, and even though it is a perturbative
contribution, the magnitude can be comparable to the background
term due to observer motion, because of the $1/|v|$ enhancement.
Let us first find the magnitude of the intrinsic term
and then the term due to motion with respect to the $n$ frame,
which gives the parallax distance.

\subsection{Perturbed parallax}

\para{The CMB rest frame.}

The vector $P_A$ is determined with respect to the $n$
frame, implying a split of the observer velocity
$u$ into the velocity of the frame of statistical homogeneity
and isotropy $n$ and the velocity $v$ with respect to it.
The reason for the split is to make more transparent
the component of the parallax that is due to our motion with
respect to distant sources, which are on average at rest with
respect to the frame of statistical homogeneity and isotropy.
The total parallax is of course is independent of the way we decompose $u$.
The frame of statistical homogeneity and isotropy is the one
in which the matter distribution is statistically homogeneous
and isotropic. If primordial perturbations are adiabatic (as
they dominantly are \cite{Planckinflation}), then this is the same
as the frame in which the CMB is statistically isotropic.
As the CMB is observed more precisely and analysed more
model-independently than galaxy catalogs, it provides a more
convenient reference point. The velocity of observers in general
differs from the velocity of the frame of statistical homogeneity
and isotropy, due to local deviations from exact homogeneity
and isotropy, but these variations vanish on average.

The CMB rest frame (also called just the CMB frame) is
the frame where the CMB dipole vanishes. As there is a subdominant
contribution to the dipole from primordial perturbations and
propagation from the last scattering surface to the observer,
this frame is different from the frame where the CMB is statistically
homogeneous and isotropic. The latter frame is sometimes
called the average CMB frame \cite{Aghanim:2013}.
The frame of statistical isotropy of the CMB can be determined
from modulation and aberration of the CMB, though at the moment
the constraints are not very precise \cite{Aghanim:2013}.
Based on inflation, the difference between the frame of statistical
homogeneity and isotropy of matter and the CMB rest frame is expected
to be at the level of 1\%, and the vanishing dipole is
a simpler condition. We therefore neglect this difference, and
take $n$ to be the frame of zero CMB dipole.

\para{Photon momentum and redshift.}

Photon momentum from source $A$ is
\bea \label{kpert}
  k_A^\a &=& \bar k_A^\a + \delta k_A^\a = a \bar k^0_{A} \bar \kh_A^\a + \delta k_A^\a \ ,
\eea

\noindent where we have $\bar k_A^0=a^{-2}\bar k^0_{A0}$ and
$\bar\kh_A^\a=(a^{-1},\be_A^i)\equiv a^{-1}(1,\te_A^i)$, so that
$\gamma_{ij}\te_A^i\te_A^j=1$.
For $n$ and $v$ we have, taking into account $n\cdot n=-1$,
\bea \label{npert}
  n^\a &=& \bar n^\a + \delta n^\a = a^{-1}\delta^{\a0} + \delta n^\a \simeq ( a^{-1} [1 - \phi] , n^i ) \equiv a^{-1} ( 1 - \phi , \tilde n^i ) \el
  v^\a &=& \delta v^\a \simeq ( 0 , u^i - n^i ) \ .
\eea

We can solve $k_A$ from the null geodesic equation $k_A^\a\nabla_\a k_A=0$.
For the time component, we get
\bea \label{deltak0}
  \frac{\delta k_A^0}{\bar k_A^0} &\simeq& - 2 \phi + B_i \te_A^i + \int^\eta \rmd\tilde\eta \left( \phi' - B'_i \te_A^i - C'_{ij} \te_A^i \te_A^j \right) \ ,
\eea

\noindent where we have dropped an integration constant,
the integral is along a background null geodesic and
prime denotes derivative with respect to $\eta$.
We see that $\delta k_A^0/\bar k_A^0\sim\sO(\e)$.

Redshift in the $n$ frame is given by $1+z_A=E^{(n)}_A/E^{(n)}_{A0}$,
so we have, from \re{E}, \re{kpert}, \re{npert} and \re{deltak0},
\bea \label{z}
  \!\!\!\!\!\! 1 + z_A &\simeq& \frac{a_0}{a} \left[ 1 - \phi + \phi_0 - \gamma_{ij} \tilde n^i \te_A^j + \gamma_{ij} \tilde n^i \te_A^j|_0 - \int_{\eta}^{\eta_0}\rmd\tilde\eta \left( \phi' - B'_i \te_A^i - C'_{ij} \te_A^i \te_A^j \right) \right] \ .
\eea

\noindent At first sight, it might be tempting to identify the
$\phi$ terms with intrinsic primordial dipole,
$\gamma_{ij}\tilde n^i\te_A^j$ terms
with kinetic dipole and the integral with dipole due to the effects
of propagation. This would amount to identifying the
hypersurface of statistical homogeneity and isotropy with the
hypersurface of constant background time, which is however a gauge
dependent quantity. This is clear if we consider the synchronous
comoving gauge\footnote{This gauge restricts the generality of the
solution. If the relation between matter and geometry is given
by the Einstein equation, it can be chosen
for irrotational dust, but not for general matter content.},
where $\phi=0, n^i=0, B_i=0$,
so that only the integral remains, but obviously the kinetic
dipole still exists (see section 5.2 of \cite{Rasanen:2011b}).
The frame of vanishing dipole is defined by
$\int\rmd\Omega e_A^\a z_A=0$, where $\int\rmd\Omega$ is an integral over the
celestial sphere at the observer, so from \re{z} we get the condition
\bea \label{dipole}
  n^k - \frac{3}{4\pi} \int\rmd\Omega e_A^k \bar g_{ij} n^i e_A^j|_\mathrm{s} = \frac{3}{4\pi} \int\rmd\Omega e_A^k \left[ \phi_\mathrm{s} + \int_{\eta_\mathrm{s}}^{\eta_0}\rmd\tilde\eta \left( \phi' - B'_i \te_A^i - C'_{ij} \te_A^i \te_A^j \right) \right] \ ,
\eea

\noindent where s refers to the source (in this case, the last scattering
surface). Excepting the possibility that $n^k$ would be correlated in
such a way that there is a large local contribution that cancels
against a large contribution from the last scattering surface, we have
$n^k\sim\sO(\e)$.

For the spatial components, it is convenient to rewrite
the null geodesic equation as
$k_A^\b\pat_\b k_{A\a}=\frac{1}{2}\pat_\a g_{\mu\nu} k_A^\mu k_A^\nu$
(as in e.g. \cite{Boubekeur:2009}) to get
\bea \label{deltaki}
  (\pat_0 + \te_A^j\pat_j) \left( \frac{\delta k_{Ai}}{\bar k_{A0}^0} \right) = - \pat_i\phi + \te_A^j \pat_i B_j + \te_A^j \te_A^k \pat_i C_{jk} + \frac{\delta k_A^j}{\bar k_A^0} \te_A^k ( \pat_i \gamma_{jk} - \pat_j \gamma_{ik} ) \ .
\eea

\noindent In the spatially flat case and for Cartesian coordinates,
we get (again dropping an integration constant) the simple result
\bea \label{deltakiflat}
  \frac{\delta k_{Ai}}{\bar k_{A0}^0} &=& \int^\eta\rmd\tilde\eta \left( - \pat_i\phi + \te_A^j \pat_i B_j + \te_A^j \te_A^k \pat_i C_{jk} \right) \el
  &=& ( - \phi + B_j \te_A^j + C_{jk} \te_A^j \te_A^k ) \be^i  \el
  && + \int^\eta\rmd\tilde\eta \left( - \nabla_{A\perp i} \phi + \phi' + \te_A^j \nabla_{\perp i} B_j - B_j' \te_A^j \te_A^i + \te_A^j \te_A^k \nabla_{\perp i} C_{jk} - C'_{jk} \te_A^j \te_A^k \te_A^i \right) \el
  &\simeq& \int^\eta\rmd\tilde\eta \left( - \nabla_{A\perp i} \phi + \te_A^j \nabla_{\perp i} B_j + \te_A^j \te_A^k \nabla_{\perp i} C_{jk} \right) \ ,
\eea

\noindent where
$\nabla_{A\perp i} \equiv ( \delta^{ij} - \te_A^i \te_A^j ) \pat_j$,
and in the last equality we have dropped all terms that are definitely
not larger than $\sO(\e)$.
The magnitude of the remaining terms could be
$\delta k_{Ai}/\bar k_{A}^0\sim\sO(\pat\e)$.
However, the integral over the derivatives can lead to
cancellations that reduce the amplitude.

\para{Intrinsic contribution.}

For $P_{\perp A}$ we get
\bea \label{Pperp}
  P_{\perp A}^0 &\simeq& 0 \el
  - \frac{1}{2} a^2 P_{\perp A}^i &\simeq& ( \delta k_A^i/\bar{k}_A^0 )' + \cH \tilde n^i + \bar\Gamma^i_{jk} \tilde n^j \te_A^k + B^{i'} + \cH B^i + \gamma^{ij} \phi_{,j} + \gamma^{ij} ( C'_{jk} + B_{[j,k]} ) \te_A^k \el
  && - \left[ (\delta k_A^0/\bar{k}_A^0 )' + \cH \gamma_{jk} \tilde n^j \te_A^k + \phi' + ( \phi_{,j} + \cH B_j ) \te_A^j \right] \te_A^i \ ,
\eea

\noindent where $\cH\equiv a'/a$. Corrections to \re{Pperp}
are $\sO(H\pat\e\pat\e)$. The intrinsic term in \re{gdot}
and \re{phiperxtime} is thus
\bea \label{intr}
  && - \frac{1}{2 |v|} ( P_{\perp 1} \cdot r_2 + P_{\perp 2} \cdot r_1 ) \el
  &\simeq& \frac{1}{|v|} \left[ ( \delta k_1^i/\bar{k}_1^0 )' + \cH \tilde n^i + \bar\Gamma^i_{jk} \tilde n^j \te_A^k + B^{i'} + \cH B^i + \gamma^{ij} \phi_{,j} + \gamma^{ij} C'_{jk} \te_1^k \right] \gamma_{il} r_2^l \el
  && - \left[ (\delta k_1^0/\bar{k}_1^0 )' + \cH \gamma_{jk} \tilde n^j \te_1^k + \phi' + ( \phi_{,j} + \cH B_j ) \te_1^j \right] a^{-1} r_1 \cdot r_2 + ( 1 \leftrightarrow 2 ) \el
  &\simeq& \frac{1}{|v|} \left[ ( \delta k_1^i/\bar{k}_1^{0} )' \gamma_{ij} r_2^j + ( 1 - r_1\cdot r_2 ) \phi_{,j} r_1^j \right] + ( 1 \leftrightarrow 2 ) \ ,
\eea

\noindent where in the second equality we have dropped subleading
terms, taking into account $n^i\sim\sO(\e)$.
The remaining terms are potentially of the same order of magnitude as
background quantities. Let us now calculate the contribution to the
parallax due to observer motion and see how it compares to \re{intr}.

\para{Motion contribution.}

The vector $P_{\perp A}$ involves, to first order, derivatives
of $\delta k_A$ with respect to the background time only.
In contrast, $\thetat_A$ and $\sigmat_{A\a\b}$ also involve
spatial derivatives of $\delta k_A$. Given
$\delta k_A^0/\bar k^0\sim\sO(\e), \delta k_A^i/\bar k^0\sim\sO(\pat\e)$,
perturbative corrections to motion parallax in \re{gdot}
and \re{phiperxtime} are potentially $\sO(H\pat^2\e)$, so they
could be as large as the background contribution.

The corrections to $\thetat$ (and thus to $d_P$) and the null shear
due to first order perturbations were calculated in
\cite{Kasai:1988}\footnote{The contribution of the null shear
in \cite{Kasai:1988} corresponds to \re{gdot} if we take $-r_2=v$.
For a general $r_2$, the calculation is more involved.},
under the assumption that they are small. Evaluating the corrections
without this assumption would require numerically integrating the
Sachs equations \re{Sachs}. Although the source term for the null shear
involves the Weyl tensor, which is $\sO(H^2\pat^2\e)$, the null
shear is known to be small in many models of structure
\cite{Bolejko:2008, Weyl, Szybka:2010, Lavinto:2013}
as well as in the real universe, as determined from observations
of image deformation \cite{Munshi:2006, Clarkson:2011c}.
The reason is that even though the Weyl tensor is locally large,
positive and negative contributions along the null geodesic can cancel
\cite{Bolejko:2008, Rasanen:2009b}.
The corrections to the parallax due to null shear
are thus subdominant to the background contribution.

The source term for $\thetat_A$ involves the Ricci tensor instead of
the Weyl tensor. Keeping only the leading terms, we have
\bea \label{motion}
  \frac{1}{2} {E_A^{(n)}}^{-1} \thetat_A &\simeq& \frac{1}{2} {{\bar{E}}_A^{(n)^{-1}}} \bar\thetat_A + \frac{1}{2} {{\bar{E}}_A^{(n)^{-1}}} \pat_i\delta k^i_A \el
  &\simeq& \bar D_{PA}^{-1} + \frac{1}{2} \pat_i \int_{\eta_\mathrm{s}}^\eta\rmd\tilde\eta \left( - \nabla_{A\perp i} \phi + \te_A^j \nabla_{\perp i} B_j + \te_A^j \te_A^k \nabla_{A\perp i} C_{jk} \right) \ ,
\eea

\noindent where the second equality holds in the spatially flat case
\re{deltakiflat}. The correction could be $\sO(H\pat^2\e)$,
\ie of the same order of magnitude as the background, or even larger.
However, the result depends on the distribution of perturbations,
not just on their amplitude. For a statistically homogeneous and
isotropic distribution, the magnitude is expected to be suppressed
by cancellations in the integral along the null geodesic
\cite{Rasanen:2008b, Rasanen:2009b}.

\para{Total parallax.}

For the total parallax we get, from \re{gdot}, \re{phiperxtime}, \re{intr}
and \re{motion}, dropping subleading terms,
\bea \label{total}
  \frac{\delta\varphi}{\delta x} &\simeq& \frac{1}{ \sqrt{ 1- ( r_1\cdot r_2 )^2 }} \left[ \frac{1}{|v|} \left( ( \delta k_1^i/\bar{k}_1^{0} )' \gamma_{ij} r_2^j + ( 1 - r_1\cdot r_2 ) \phi_{,j} r_1^j \right) \right. \el
  && + \left. ( r_2\cdot \vh - r_1\cdot r_2 \ r_1\cdot \vh ) \left( \bar D_{P1}^{-1} + \frac{1}{2} \pat_i ( \delta k^i_1 / \bar k_1^0 ) \right) \right] + ( 1 \leftrightarrow 2 ) \ ,
\eea

\noindent where all terms are again evaluated at the observer, and
in the spatially flat case $\delta k^i_A$ is given by \re{deltakiflat}.
We have assumed that the observer motion is geodesic, so we can solve
$u^i$ from the condition $\udot^i=0$, with the result
\bea
  u^i &\simeq& a^{-2} A^i - a^{-1} B^i - a^{-2} \int^t \rmd\tilde t \gamma^{ij} \pat_j\phi \ ,
\eea

\noindent where $A^i$ does not depend on $t$ and is related to the initial conditions.
If $B^i=0$, we neglect $A^i$ and assume that we can write $\phi=b(t) f(x^i)$
(as is the case in linear theory when the decaying mode can be dropped)\footnote{As
mentioned in \sec{sec:geom}, linear perturbation theory cannot correctly
describe an observer velocity corresponding to motion around the Solar
system or the center of the Galaxy: the time evolution of $u^i$
(and thus $v^i$) is the same everywhere, no complicated local motions are possible.},
taking into account $v^i\simeq u^i$ we get
$|v|\simeq \frac{\int^t\rmd\tilde t b}{a b} \sqrt{\gamma^{jk}\pat_j\phi\pat_k\phi}$
and $\vh^i=-\frac{a^{-1}\gamma^{ij}\pat_j\phi}{\sqrt{\gamma^{kl}\pat_k\phi\pat_l\phi}}$.
If we also assume that the $\delta k^i_A$ terms are subdominant, \re{total} reduces to
\bea \label{total2}
  \frac{\delta\varphi}{\delta x} &\simeq& \frac{1}{ \sqrt{ 1 - ( r_1\cdot r_2 )^2 }} ( r_2\cdot \vh - r_1\cdot r_2 \ r_1\cdot \vh ) \left( \bar D_{P1}^{-1} - \frac{b_0}{\int^{t_0}\rmd t b} \right) + ( 1 \leftrightarrow 2 ) \ .
\eea

\noindent The intrinsic parallax is of the same order as the
motion parallax. It contributes in the same way as the local
expansion rate in a FRW universe with spacelike separation between observation
points and an observer whose frame is expanding with the background,
discussed in \re{sec:spaceFRW}, but with a different numerical coefficient.
For example, if we assume that the Einstein equation holds, the matter is
dust and the spatial curvature is zero, we have $b(t)=1$ and
$\frac{b_0}{\int^{t_0}\rmd t b}=t_0^{-1}=\frac{3}{2} H_0$, compared to $H_0$
for the expanding baseline case. In principle, it would be possible to
measure the time-dependence of the gravitational potential from the intrinsic
parallax, but as it is sensitive to the local gravitational potential,
the linear theory prediction cannot be trusted.

\para{Smallness of the parallax.}

The reason that the intrinsic term and the motion term are of the same
order of magnitude in \re{total2} is that the observer's deviation from
the frame of homogeneity  and isotropy is generated by the same
perturbations that cause the intrinsic parallax  term, and both are
$\sO(\pat\e)$. With the parallax distance defined as
$D_{PA}^{-1}=\frac{1}{2} {E_A^{(n)}}^{-1} \thetat_A$,
determining $D_{PA}$ from parallax measurements requires disentangling
the intrinsic term and the motion term, not just measuring the
amplitude of the parallax.

In \sec{sec:intro} we noted that in the FRW universe, both CMB anisotropy
and parallax vanish for observers comoving with the frame of  homogeneity
and isotropy.
The CMB result is stable in the direction that if the spacetime is close to FRW,
the anisotropy is small. However, small CMB anisotropy does not imply
that the spacetime is close to FRW
\cite{almostEGS, moreEGS, Rasanen:2009a, Maartens:2011}.
The parallax is a dimensional quantity, so it has to be compared
to some scale (this is also true when discussing the CMB in
the non-FRW case, as derivatives of the temperature anisotropy are
involved \cite{almostEGS, Rasanen:2009a, Maartens:2011}). 
In the cosmological case, the relevant scale is given by the Hubble
parameter, and for time intervals much smaller than the Hubble time, 
the parallax is small, $\delta\varphi\sim H_0|v|\delta\tau$.
The reverse result would mean that small parallax implies that the
universe is close to FRW (or, more generally, close to being
conformally stationary). This would mean that the terms in
\re{gdot} are always of the order of the Hubble scale or larger
if the spacetime is not close to FRW, which seems unlikely.
Indeed, studies of parallax in models with large inhomogeneities or
anisotropies
\cite{Quercellini:2008, Quartin:2009, Quercellini:2010, Amendola:2013, anisopara}
provide counterexamples (though studies of spherical
inhomogeneities have concentrated on locations close to the center,
and it is not obvious that the metrics considered cannot be written
in terms of a linearly perturbed FRW metric).
Note that the smallness of $P_{A\perp}$ means that $n^\a\nabla_\a k_A$
is almost parallel to $k_A$; it would be interesting to study this
constraint further.

Closeness to the FRW case can also be considered in terms of the FRW
consistency condition introduced in \sec{sec:con}. The magnitude of the
corrections to the consistency condition depends on the integral term
appearing in $\delta k_A^i$, see \re{deltaki} and \re{deltakiflat}.
If it is much smaller than background quantities, the FRW
consistency condition holds well.

\section{Discussion} \label{sec:disc}

\para{Non-perturbative deviations from the FRW case.}

We have only discussed perturbative deviations from the FRW case.
However, even if the universe is statistically homogeneous and isotropic,
it is possible that the mean large-scale optical properties of
the universe are not well described by the FRW metric, or that
the evolution of the average expansion rate over large volumes
is not well described by the FRW equations
\cite{Ellis:2005, Rasanen:2011a, Buchert:2011, Rasanen:2006b}.
In either case, the metric is not perturbatively close to
a single FRW metric everywhere \cite{Rasanen:2011b}.
Another issue that has to be taken into account is that even if the
metric at our location is perturbatively close to the FRW metric, it does not
satisfy the linearly perturbed FRW equations, as discussed in \sec{sec:geom}
\cite{Rasanen:2010a}.
So a description beyond linear order, and perhaps beyond perturbation
theory, has to be adopted to properly describe observers
inside stabilised regions, particularly in the case of complex local
motion, such as orbiting the Sun and the center of the Galaxy.

The first study to consider, in connection with the parallax,
the possibility that the mean optical properties of the
universe are not described by the FRW model was done in
the 1970s, using the Dyer--Roeder prescription
(which was developed by Zel'dovich) \cite{Novikov:1977, Novikov:1978}.
In the Dyer--Roeder approach, the spacetime dynamics are assumed to
be the same as in the FRW model, but the energy density along the null
geodesics is taken to be smaller than the FRW density.
However, the expansion rate along the null geodesics is taken to be
identical to the one in the FRW model. This is inconsistent, and
the expansion rate affects the distance via the mapping of the affine
parameter to the redshift \cite{Rasanen:2008b, Rasanen:2009b, Clarkson:2011c}.
It has been argued that for a statistically homogeneous and isotropic
distribution the relation between the average expansion rate and the
affine parameter is the same as in the FRW model,
even though the average expansion rate may be different from the FRW case
\cite{Rasanen:2008b, Rasanen:2009b, Bull:2012, Lavinto:2013}.
It is also expected that $\thetat$, which gives the parallax
distance, is related to the average expansion rate and average density
\cite{Rasanen:2008b}.
In this case, the consistency condition \re{kP} between $d_P$ and $d$
reduces to testing the relation \re{dFRW} between $h$ and $d$:
if $k_H$ is constant, this implies that $k_P$ is also constant
and equal to $k_P$. In general, if the average expansion rate is
not close to the FRW case, $k_H$ and $k_P$ are expected to be
different and vary with redshift \cite{Boehm:2013, Lavinto:2013}.

\para{Testing the FRW metric vs. testing homogeneity and isotropy.}

The FRW consistency condition \re{kP} introduced in \sec{sec:con} is a
simple but powerful test. If $d$ and $d_P$ are measured at any two
redshifts (that are sufficiently large compared to the homogeneity
scale) and the corresponding values of $k_P$ are not equal, the
universe on large scales is not described by the four-dimensional FRW
metric. More precisely, \re{kP} tests the optical properties of the
universe, no assumptions have been made about the matter content or
its relation to the spacetime geometry.  The only assumptions are that
the spacetime manifold is 3+1-dimensional and spatially homogeneous
and isotropic, and that light propagation can be approximated by
geometrical optics. The consistency condition \re{kP} tests
exact homogeneity and isotropy. We have shown that, apart possibly
from the integral term in \re{deltaki} and \re{deltakiflat},
this test is stable to linear perturbations.
However, if the mean optical properties of the universe
are not well described by the FRW metric, the consistency condition
will in general be violated, even if the universe is statistically
homogeneous and isotropic.

Several tests of homogeneity and isotropy have been proposed
\cite{Clarkson:2007b, Maartens:2011, stathom, homrev, Clifton:2011}.
Many of them test statistical homogeneity and isotropy,
though this is sometimes conflated with exact homogeneity and isotropy
\cite{stathom}. Others test for specific deviations from homogeneity
and isotropy, such as large spherical inhomogeneities, and are thus
particular tests of the Copernican principle (which is a distinct
issue from homogeneity and isotropy; see section 1.3 of \cite{Buchert:2011}).

Some proposals test for exact homogeneity and isotropy
\cite{EGS, Clifton:2011}. However, the universe is not described by the
exact FRW metric, and there are locally deviations of order unity in
geometrical quantities such as the expansion rate. The relevant
questions are then whether the universe is statistically
homogeneous and isotropic and whether the metric is perturbatively
close to the same FRW model everywhere.
These are independent issues. On the one hand,
statistical  homogeneity and isotropy does not imply that the universe
is perturbatively close to FRW, or that light propagation or spatial
expansion is well described by the FRW model \cite{Lavinto:2013}.
On the other hand, without statistical homogeneity, the metric being close
to FRW is not a sufficient condition for observables such as the
distance $d$ to be close to the FRW case \cite{Enqvist:2009}. 

Tests of the exact FRW metric are useful if they can be
extended to the near-FRW situation. Then they will test
either statistical homogeneity and isotropy (like CMB near-anisotropy
\cite{Rasanen:2009a}), or being perturbatively close
to the FRW metric. The former are useful for testing models with large
spherical inhomogeneities or models with significant global anisotropy.
In principle, such tests could also be used to constrain
large-scale isocurvature perturbations by comparing the frame of
homogeneity and isotropy as inferred from the CMB and from light
rays sourced by the matter distribution, though in practice the
precision is likely to remain lower than with other methods.
At the moment, the only tests that probe whether the universe
is close to FRW are the consistency condition \re{kP} and the
consistency condition \re{kH} proposed in \cite{Clarkson:2007b}.
These two consistency conditions have the virtue
of using only the metric and geometrical optics.
Other such kinematic geometrical optics tests of the
FRW metric can be devised and observationally tested.

\para{Measuring cosmic parallax with Gaia.}

The Gaia satellite, launched on December 19, 2013,
will probe parallax at cosmological distances.
The precision for a single source depends on the
magnitude, but for galaxies and quasars at cosmological distances,
it is typically of the order 100 $\mu$as
\cite{Mignard:2005, Lindegren:2008, Ding:2009, Quartin:2009, Quercellini:2010}.
Assuming that deviations from the FRW model are at most of order unity,
the order of magnitude of cosmic parallax is
$\delta\varphi\sim H_0 \delta x\sim H_0 |v| \delta\tau\sim 10^{-2} \mu$as $\delta\tau/$yr $\sim10^{-1} \mu$as
for $|v|=$ 369 km/s \cite{Aghanim:2013} and the Gaia mission
duration $\delta\tau=$ 5 yr.
The precision for an individual source is therefore three orders of
magnitude smaller than the expected cosmic parallax, and
parallax distance to individual objects can be measured
only for distances of up to about Mpc.

However, Gaia is expected to measure $N=5\times 10^5$ quasars, and
measurement error goes down by a factor of $1/\sqrt{2N}\sim10^{-3}$,
bringing the precision close to the expected cosmic signal
\cite{Ding:2009, Quartin:2009, Quercellini:2010}.
It is not clear how well it will be possible to measure the redshift
dependence of $d_P$, as the quasars are
spread over a large range of redshifts up to $z=5$, the number
peaking between $z=1$ and $z=2$ \cite{Slezak:2007, Robin:2012, Gaiasp}.
As noted in \sec{sec:con}, $d_P(z)$ saturates at large $z$.
For example, in the spatially flat \LCDM model with $\Omn=0.3$ we have
$d_P(5)/d_P(1)\approx1.5$, so parallax remains at roughly the
same order of magnitude for all large redshifts probed by Gaia, and
increase in measurement difficulty is mostly due to decreasing luminosity.
See \cite{Ding:2009, Quartin:2009} for the expected precision as a function
of redshift, taking the magnitude distribution of quasars into account.
In addition to quasars, Gaia is expected to measure the redshifts of
$3\times 10^6$ galaxies up to $z=0.75$ \cite{Robin:2012}, which have
not been included in analysis of cosmic parallax so far.

For testing the consistency condition \re{kP}, only those redshifts are useful
for which there are also measurements of $d$. Current observations of type
Ia supernovae measure $d_L=(1+z) d$ only up to $z=1.4$ \cite{Suzuki:2011},
and $d_A=(1+z)^{-1} d$ has been measured up to $z=2.4$ using baryon acoustic
oscillations \cite{Slosar:2013}, though
the analysis is more model-dependent than in the case of supernovae.
As the test \re{kP} does not involve derivatives or integrals, the measurement
of $d$ needs to be less precise than for the consistency condition
\re{kH} between $d$ and $h$, though measuring
$d_P$ is more difficult than measuring $h$.
Constraints on $k_H$ are at the moment of order
$|k_H|\lesssim1$ \cite{Shafieloo:2009, Mortsell:2011},
though they are likely to improve perhaps by an order of magnitude
when all current and near-future data is included in the analysis.

In Gaia analyses, some faraway sources are assumed to have no
parallax (except due to our motion around the center of the Galaxy)
and are used to establish a cosmic rest frame
\cite{Ding:2009, Quartin:2009, Mignard:2005, Slezak:2007, Robin:2012}.
This is potentially problematic from the point of view of measuring
cosmic parallax, because in a perturbed FRW spacetime the motion parallax
of all sources at cosmological distances is of the same order of magnitude,
and this is also the order of magnitude of the intrinsic parallax.

We have not considered small-scale motion around the Sun and the
center of the Galaxy.
In the literature, the effect of the small-scale velocity
and acceleration on the parallax has been studied with Newtonian physics
\cite{Kovalevsky:2003, Mignard:2005, Slezak:2007, Liu:2013},
but it would be interesting to include it in a consistent covariant
relativistic treatment.
Treatment of the local environment is particular important, as the
intrinsic parallax depends on quantities at the location of the observer.
The motion of the sources with respect to the
frame of statistical homogeneity and isotropy should also be taken into
account. As measurements on cosmological distances depend on a
large number of sources, this is expected to only add random noise,
which can be distinguished from the cosmological signal \cite{Ding:2009, Quartin:2009}.

The actual precision with which Gaia will measure, or put limits on,
the cosmic parallax will depend on the real source distribution
(as opposed to the simulated catalogue) and the actual performance
of the instrument. The theoretical analysis also needs to be extended to
address issues discussed here. Nevertheless, it seems feasible that Gaia could
provide a measurement of $d_P$ at cosmological distances with an accuracy
of at least order unity, leading to a test of the FRW metric at a
precision comparable to present tests of $k_H$.

\section{Conclusion} \label{sec:conc}

\para{Summary and outlook.}

The Gaia satellite offers the possibility of measuring parallax over
cosmological distances for the first time in the next few years.
Classic discussions of cosmic parallax have considered spacelike
separation between observation points and a single source.
Using the covariant formalism and considering the angle between
a pair of sources, we have calculated the parallax for
both spacelike and timelike separation between observation points.
We have included both the intrinsic parallax due to the fact that
the spacetime is not conformally stationary and the parallax due to
observer motion with respect to the mean rest frame of distant sources
(which for adiabatic initial conditions is close to the CMB rest frame).

In principle, parallax offers a way to measure cosmological
distances without any assumptions about source properties,
unlike the angular diameter distance $d_A$ or the luminosity
distance $d_L$.
In practice, parallax is so small that over cosmological distances
it cannot in the near future compete with measurements
of angular diameter or luminosity.
However, because the parallax distance contains independent information
from $d_A$ (and $d_L$), their combination makes it possible to test
whether the large-scale optical properties
of the universe are described by the four-dimensional FRW metric.
We have turned this possibility, first mentioned in \cite{Mccrea:1935},
into a concrete proposal.
The test is independent of dynamical issues related to matter content and its
relation to the spacetime geometry, it depends only on the validity of
the FRW metric and geometrical optics. Such kinematic geometrical
optics tests are simple but powerful: a negative result would rule out all
four-dimensional FRW models where light propagates as in general relativity.
A similar consistency condition 
between the expansion rate and $d_A$ (or $d_L$) was introduced in
\cite{Clarkson:2007b}, and has been tested observationally
\cite{Shafieloo:2009, Mortsell:2011}.  Observations of the Gaia
satellite may make it possible to test the parallax
consistency condition for the first time.

We have studied the stability of the parallax due to our motion with
respect to the frame of homogeneity and isotropy to linear
perturbations around the FRW metric. There is one integral term
that may potentially be of the same order of magnitude as the background
contribution, and the intrinsic parallax due to the
perturbations is of the same order of magnitude as the parallax due
to our motion. Therefore, when analysing observations, perturbations have to
be taken into account. The intrinsic term may be distinguished by its
different dependence on redshift.

The perturbative integral correction to the parallax should be evaluated.
It should also be taken into account that the local environment of observers
located in a stabilised region and undergoing orbital motions cannot
be described
by the linear equations, but has to be considered with non-linear perturbation
theory, or a non-perturbative treatment. The capability of the Gaia satellite to
measure cosmic parallax should be analysed in more detail, possibly by
combining galaxy and quasar data. In particular, the procedure that some quasars
are taken to define a cosmic rest frame by having vanishing intrinsic parallax
should be carefully considered.

In addition to the consistency condition between parallax distance
and $d_A$ (or $d_L$) proposed here and the condition between the
expansion rate and $d_A$ (or $d_L$) proposed in \cite{Clarkson:2007b},
it is possible to develop other kinematic tests of the FRW metric
based on geometrical optics.

\acknowledgments

I thank Stanislav Rusak for translating \cite{Kardashev:1973}.

\end{document}